\documentclass{aa}
\usepackage{psfig_laa}
\begin{document}
\thesaurus{08(09.03.1,09.13.2,09.19.1,10.03.1,13.19.3)}%
%
%
%
\def\etal {et al.}
\def\ie {i.\,e.}
\def\etseq {{\em et seq.}}
\def\vs {{it vs.}}
\def\perse {{it per se}}
\def\adhoc {{\em ad hoc}}
\def\eg {e.\,g.}
\def\etc {etc.}
\def\ccpers {\hbox{${\rm cm}^3{\rm s}^{-1}$}}
\def\DEGR {\hbox{$^{\circ }$}}
\def\vlsr {\hbox{${v_{\rm LSR}}$}}
\def\vel {\hbox{${v_{\rm LSR}}$}}
\def\vhel {\hbox{${v_{\rm HEL}}$}}
\def\delv {\hbox{$\Delta v_{1/2}$}}
\def\dvel {\hbox{$\Delta v_{1/2}$}}
\def\TL {$T_{\rm L}$}
\def\TC {$T_{\rm c}$}
\def\TEX {$T_{\rm ex}$}
\def\TMB {$T_{\rm MB}$}
\def\TKIN {$T_{\rm kin}$}
\def\TREC {$T_{\rm rec}$}
\def\TSYS {$T_{\rm sys}$}
\def\TVIB {$T_{\rm vib}$}
\def\TROT {$T_{\rm rot}$}
\def\TDUST {$T_{\rm d}$}
\def\TASTAR {$T_{\rm A}^{*}$}
\def\TVIBST {$T_{\rm vib}^*$} 
\def\TB {$T_{\rm B}$}
\def \la{\mathrel{\mathchoice   {\vcenter{\offinterlineskip\halign{\hfil
$\displaystyle##$\hfil\cr<\cr\sim\cr}}}
{\vcenter{\offinterlineskip\halign{\hfil$\textstyle##$\hfil\cr
<\cr\sim\cr}}}
{\vcenter{\offinterlineskip\halign{\hfil$\scriptstyle##$\hfil\cr
<\cr\sim\cr}}}
{\vcenter{\offinterlineskip\halign{\hfil$\scriptscriptstyle##$\hfil\cr
<\cr\sim\cr}}}}}
\def \ga{\mathrel{\mathchoice   {\vcenter{\offinterlineskip\halign{\hfil
$\displaystyle##$\hfil\cr>\cr\sim\cr}}}
{\vcenter{\offinterlineskip\halign{\hfil$\textstyle##$\hfil\cr
>\cr\sim\cr}}}
{\vcenter{\offinterlineskip\halign{\hfil$\scriptstyle##$\hfil\cr
>\cr\sim\cr}}}
{\vcenter{\offinterlineskip\halign{\hfil$\scriptscriptstyle##$\hfil\cr
>\cr\sim\cr}}}}}
\def\RZWCO {${\cal R}_{2/1}^{\rm C^{18}O}$}
\def\RDRCO {${\cal R}_{3/2}^{\rm C^{18}O}$}
\def\RMSIO {${\cal R}_{5/2}^{\rm ^{28}SiO}$}
\def\RISOSIO {${ r }_{28/29}^{\rm SiO}$}
\def\RISOZSIO {${ r}_{29/30}^{\rm SiO}$}
\def\H0 {$H_{\rm o}$}
\def\mic {$\mu\hbox{m}$}
\def\micro {\mu\hbox{m}}
\def\SDOZ {\hbox{$S_{12\mu \rm m}$}}
\def\STWE {\hbox{$S_{25\mu \rm m}$}}
\def\SSIX {\hbox{$S_{60\mu \rm m}$}}
\def\SHUN {\hbox{$S_{100\mu \rm m}$}}
\def\solmass {\hbox{M$_{\odot}$}}
\def\solum {\hbox{L$_{\odot}$}}
\def\irlum {\hbox{$L_{\rm IR}$}}
\def\ohlum {\hbox{$L_{\rm OH}$}}
\def\blum {\hbox{$L_{\rm B}$}}
\def\numd {\hbox{$n\,({\rm H}_2$)}}                   
\def\rhounit {$\hbox{M}_\odot\,\hbox{pc}^{-3}$}
\def\kms {\hbox{${\rm km\,s}^{-1}$}}
\def\kmsyr {\hbox{${\rm km\,s}^{-1}\,{\rm yr}^{-1}$}}
\def\kmsmpc {\hbox{${\rm km\,s}^{-1}\,{\rm Mpc}^{-1}$}} 
\def\Kkms {\hbox{${\rm K\,km\,s}^{-1}$}}
\def\percc {$\hbox{{\rm cm}}^{-3}$}    
\def\cmsq  {$\hbox{{\rm cm}}^{-2}$}    
\def\cmsix  {$\hbox{{\rm cm}}^{-6}$}  
\def\arcsec {\hbox{$^{\prime\prime}$}}
\def\arcmin {\hbox{$^{\prime}$}}
\def\ffam {\hbox{$\,.\!\!^{\prime}$}}
\def\ffas {\hbox{$\,.\!\!^{\prime\prime}$}}
\def\ffM {\hbox{$\,.\!\!\!^{\rm M}$}}
\def\ffm {\hbox{$\,.\!\!\!^{\rm m}$}}
\def\ffd {\hbox{$\,.\!\!^{\circ}$}}
\def\HI  {\hbox{HI}}
\def\HII {\hbox{HII}}
%
%
\def \AL {$\alpha $}    
\def \BE {$\beta $}     
\def \GA {$\gamma $}    
\def \DE {$\delta $}    
\def \EP {$\epsilon $}  
\def \alde {($\Delta \alpha ,\Delta \delta $)}
\def \MU {$\mu $}       
\def \TAU {$\tau $}     
\def \tapp {$\tau _{\rm app}$}
\def \tuns {$\tau _{\rm uns}$}
\def \RH {\hbox{$R_{\rm H}$}}         
\def \RT {\hbox{$R_{\rm \tau}$}}      
\def \BN  {\hbox{$b_{\rm n}$}}        
\def \BETAN {\hbox{$\beta _n$}}       
\def \TE {\hbox{$T_{\rm e}$}}         
\def \NE {\hbox{$N_{\rm e}$}}         
%
\def\MOLH {\hbox{${\rm H}_2$}}                    
\def\HDO {\hbox{${\rm HDO}$}}                     
\def\AMM {\hbox{${\rm NH}_{3}$}}                  
\def\NHTWD {\hbox{${\rm NH}_2{\rm D}$}}           
\def\CTWH {\hbox{${\rm C_{2}H}$}}                 
\def\TCO {\hbox{${\rm ^{12}CO}$}}                 
\def\CEIO {\hbox{${\rm C}^{18}{\rm O}$}}          
\def\CSEO {\hbox{${\rm C}^{17}{\rm O}$}}          
\def\CTHFOS {\hbox{${\rm C}^{34}{\rm S}$}}        
\def\THCO {\hbox{$^{13}{\rm CO}$}}                
\def\WAT {\hbox{${\rm H}_2{\rm O}$}}              
\def\WATEI {\hbox{${\rm H}_2^{18}{\rm O}$}}       
\def\CYAN {\hbox{${\rm HC}_3{\rm N}$}}            
\def\CYACFI {\hbox{${\rm HC}_5{\rm N}$}}          
\def\CYACSE {\hbox{${\rm HC}_7{\rm N}$}}          
\def\CYACNI {\hbox{${\rm HC}_9{\rm N}$}}          
\def\METH {\hbox{${\rm CH}_3{\rm OH}$}}           
\def\MECN {\hbox{${\rm CH}_3{\rm CN}$}}           
\def\METAC {\hbox{${\rm CH}_3{\rm C}_2{\rm H}$}}  
\def\CH3C2H {\hbox{${\rm CH}_3{\rm C}_2{\rm H}$}} 
\def\FORM {\hbox{${\rm H}_2{\rm CO}$}}            
\def\MEFORM {\hbox{${\rm HCOOCH}_3$}}             
\def\THFO {\hbox{${\rm H}_2{\rm CS}$}}            
\def\ETHAL {\hbox{${\rm C}_2{\rm H}_5{\rm OH}$}}  
\def\CHTHOD {\hbox{${\rm CH}_3{\rm OD}$}}         
\def\CHTDOH {\hbox{${\rm CH}_2{\rm DOH}$}}        
\def\CYCP {\hbox{${\rm C}_3{\rm H}_2$}}           
\def\CTHHD {\hbox{${\rm C}_3{\rm HD}$}}           
\def\HTCN {\hbox{${\rm H^{13}CN}$}}               
\def\HNTC {\hbox{${\rm HN^{13}C}$}}               
\def\HCOP {\hbox{${\rm HCO}^+$}}                  
\def\HTCOP {\hbox{${\rm H^{13}CO}^{+}$}}          
\def\NNHP {\hbox{${\rm N}_2{\rm H}^+$}}           
\def\CHTHP {\hbox{${\rm CH}_3^+$}}                
\def\CHP {\hbox{${\rm CH}^{+}$}}                  
\def\ETHCN {\hbox{${\rm C}_2{\rm H}_5{\rm CN}$}}  
\def\DCOP {\hbox{${\rm DCO}^+$}}                  
\def\HTHP {\hbox{${\rm H}_{3}^{+}$}}              
\def\HTWDP {\hbox{${\rm H}_{2}{\rm D}^{+}$}}      
\def\CHTWDP {\hbox{${\rm CH}_{2}{\rm D}^{+}$}}    
\def\CNCHPL {\hbox{${\rm CNCH}^{+}$}}             
\def\CNCNPL {\hbox{${\rm CNCN}^{+}$}}             
%
%
\def\In {\hbox{$I^{n}(x_{\rm k},y_{\rm k},u_{\rm l}$})}
\def\Iobs {\hbox{$I_{\rm obs}(x_{\rm k},y_{\rm k},u_{\rm l})$}}
\def\Ingl {I^{n}(x_{\rm k},y_{\rm k},u_{\rm l})}
\def\Iobsgl {I_{\rm obs}(x_{\rm k},y_{\rm k},u_{\rm l})}
\def\Pbgl {P_{\rm b}(x_{\rm k},y_{\rm k}|\zeta _{\rm i},\eta _{\rm j})}
\def\Pbgm {P(x_{\rm k},y_{\rm k}|r_{\rm i},u_{\rm l})}
\def\Pbgn {P(x,y|r,u)}
\def\Pugm {P_{\rm u}(u_{\rm l}|w_{\rm ij})}
\def\Pdem {P_{\rm b}(x,y|\zeta (r,\theta ),\eta (r,\theta ))} 
\def\Pden {P_{\rm u}(u,w(r,\theta ))}
\def\greekgl {(\zeta _{\rm i},\eta _{\rm j},u_{\rm l})}
\def\greekg1 {(\zeta _{\rm i},\eta _{\rm j})}
\title{Molecular gas in the Galactic center region}
\subtitle{III. Probing shocks in molecular cores\thanks{Based on observations
obtained at the Swedish-ESO Submillimeter Telescope (SEST, Project 
C-0518, 1996), the James Clark Maxwell Telescope (JCMT, operated
by the Joint Astronomy Centre on behalf of the Particle Physics and
Astronomy Research Council of the United Kingdom, the Netherlands 
Organisation for Scientific Research, and the National Research Council of Canada) and the Heinrich-Hertz-Telescope (HHT, operated by the the 
Submillimeter Telescope Observatory).}}
\author{S.~H\"{u}ttemeister\inst{1,2}, G.~Dahmen\inst{3,4}, 
R.~Mauersberger\inst{5}, C.~Henkel\inst{4}, T.L.~Wilson\inst{4,6}, 
J.~Mart\'{\i}n-Pintado\inst{7} }
\offprints{S. H{\"u}ttemeister, RAIUB}
\institute{
 Radioastronomisches Institut der Universit\"{a}t Bonn,
 Auf dem H\"{u}gel 71, D - 53121 Bonn, Germany
\and
 Harvard-Smithsonian Center for Astrophysics,
 60 Garden Street, Cambridge, MA 02138, U.S.A.
\and
 Physics Department, Queen Mary \& Westfield College,
 University of London, Mile End Road, London E1 4NS, England
\and
 Max-Planck-Institut f{\"u}r Radioastronomie,
 Auf dem H{\"u}gel 69, D - 53121 Bonn, Germany
\and
 Steward Observatory, The University of Arizona,
 Tucson, AZ 85721, U.S.A.
\and
 Submillimeter Telescope Observatory, The University of Arizona,
 Tucson, AZ 85721, U.S.A.
\and
 Centro Astron{\'o}mico de Yebes, Apartado 148, 
 19080 Guadalajara, Spain }
\date{Received 1 December1997 / Accepted 24 February 1998 }
\titlerunning{Molecular Gas in the GC Region III.~Probing shocked molecular
gas}
\authorrunning{S.~H\"uttemeister et al.}
\maketitle
\begin{abstract}
Multiline observations of C$^{18}$O and SiO isotopomers
toward 33 molecular peaks in the Galactic center region, 
taken at the SEST, JCMT and HHT telescopes, are presented. 
The C$^{18}$O presumably traces the total H$_2$ column density, 
while the SiO traces gas affected by shocks and high temperature 
chemistry. The $J =2\to 1$ line of SiO is seen only in few regions 
of the Galactic disk. This line is easily detected in all Galactic 
center sources observed. A comparison of the strength of the rare 
isotopomers $^{29}$SiO and $^{30}$SiO to the strength of the main
isotopomer $^{28}$SiO implies that the $J = 2 \to 1$ transition 
of $^{28}$SiO is optically thick. The $^{29}$Si$/^{30}$Si isotope 
ratio of 1.6 in the Galactic center clouds is consistent 
with the terrestrial value. Large Velocity Gradient models show that 
the dense component ($n_{\rm H_2} \geq  10^4$\,\percc) in typical 
molecular cores in the Galactic center is cool (\TKIN{} $\approx 
25$\,K), contrary to what is usually found in Giant Molecular Clouds 
in the disk, where the densest cores are the hottest. High
kinetic temperatures, $> 100$\,K, known to exist from NH$_3$ 
studies, are only present at lower gas densities of a few 
10$^3$\,\percc , where SiO is highly subthermally excited. Assuming that 
\CEIO\ traces all of the molecular gas, it is found that in all cases 
but one, SiO emission is compatible with arising in gas at higher 
density that is (presently) relatively cool. The relative abundance of 
SiO is typically $10^{-9}$, but differs significantly between individual 
sources. It shows a dependence on the position of the source within 
the Galactic center region. High abundances are found in those regions
for which bar potential models predict a high likelihood for cloud-cloud
collisions. These results can be used to relate the amount of gas that 
has encountered shocks within the last $\sim 10^6$ years to the large scale 
kinematics in the inner $\sim$\,500\,pc of the Galaxy.
\end{abstract} 
\section{Introduction}
The molecular environment in the inner $\sim 8^{\circ}$ of the Galaxy 
differs drastically from that in the Galactic disk. Large scale 
surveys of CO isotopomers (e.g.\ Bally et al.\ 1987, Heiligman 1987, 
Jackson et al.\ 1996, Dahmen et al.\ 1997a (Paper\,I), Bitran et al.\ 
1997) show that the cloud and the intercloud medium in the gaseous 
bulge of the Milky Way are molecular. The gas is characterized by 
large linewidths, indicating a high degree of turbulence. As shown
by Dahmen et al.\ 1998 (Paper\,II), there is no simple relationship
between CO line intensities and H$_2$ column densities and 
thus no global $N({\rm H}_2)/I$(CO) conversion factor. 

Except for a few extraordinary regions such as Sgr\,A and Sgr\,B2, 
little evidence for ongoing massive star formation in Galactic center
Giant Molecular Clouds (GMCs) is found, as is demonstrated by a general lack
of strong FIR or radio continuum point sources associated with these
clouds (Odenwald \& Fazio 1984, G\"usten 1989). Ambient dust 
temperatures are fairly low at \TDUST\ $\sim 25 - 30$\,K (e.g.\ Cox \& Laurijs 
1989). Based on ISO data, this is also a typical \TDUST\ in the 
sources studied in this paper. 

In the Galactic disk, quasithermal SiO emission is tightly correlated with 
high temperature regions (e.g.\ Ziurys et al.\ 1989). A close association
with outflows strongly suggests that grain disruption by shocks is the
major mechanism  for releasing SiO into the gas phase (Mart\'{\i}n-Pintado 
et al.\ 1992), although high temperature gas phase chemistry (Langer 
\& Glassgold 1990) may play a minor role. In the Galactic center region, 
SiO is much more widespread (see e.g.\ the survey in the $J=1\to 0$
transition by Mart\'{\i}n-Pintado et al.\ 1997), which was interpreted
as evidence for large scale or ubiquitous (fast) shocks. 

In this paper, \CEIO\ and SiO data are presented for molecular cloud
cores of the Bally et al.\ (1987) CS survey. Our measurements allow to 
(1) trace the H$_2$ column density, (2) determine or constrain density
and temperature structure, (3) estimate SiO abundances and (4) obtain 
information about silicon isotope ratios. Since our sources are selected 
on the basis of their intensity in CS, a general high density tracer,
the sample is not a priori biased toward strong SiO emission. 

NH$_3$ data (H\"uttemeister et al.\ 1993b) show that at least two 
phases of different kinetic temperatures are present within 
all cloud cores without massive star formation:  A cool component 
with \TKIN\ of 20\,--\,30\,K, close to the temperature of the dust in 
the Galactic center and a warm component with \TKIN\,$\geq 120$\,K. 
\TKIN\,$\sim 75$\,K is considered by many authors, (e.g.\ the reviews by 
Morris \& Serabyn 1996, Mezger et al.\ 1996) as `typical'. This is,
however, just the average over the hot and cool component and has no 
meaning as a distinct physical component. 

A major aim of this study is to determine the physical parameters,
distribution and origin of these phases and to decide which of them, 
cool or warm, is associated with the bulk of the SiO emission. 

\section{Observations}

The observations of the $J = 1 \to 0$ and $2\to 1$ transitions of 
\CEIO\ and the $J = 2\to 1$ transition of $^{28}$SiO, $^{29}$SiO and 
$^{30}$SiO were carried out at the Swedish-ESO Submillimetre Telescope 
(SEST) at La Silla (Chile). All C$^{18}$O($3 \to 2$) data were taken 
at the James Clark Maxwell Telescope (JCMT) on Mauna Kea, (Hawaii), 
where we also obtained some $^{28}$SiO($5 \to 4$) and \CEIO 
($2 \to 1$) measurements for calibration purposes. Finally, the 
$5 \to 4$ transition of $^{28}$SiO was measured at the Heinrich Hertz 
Telescope (HHT) on Mt.\ Graham (Arizona). Table \ref{log} summarizes 
the observations, including line frequencies, telescope parameters, 
receivers and spectrometers used and typical system temperatures. 
A telescope beamsize of 46$''$ corresponds to 1.9\,pc at a distance of 
the Galactic center of 8.5\,kpc.

\begin{table*}
\begin{flushleft}
\caption{\label{log} Summary of the observations }
\begin{tabular}{lrrrrrr|rrr|r}
Line & \multicolumn{1}{c}{Frequency} & \multicolumn{1}{c}{Telescope} & 
\multicolumn{1}{c}{Receiver$^{\rm a)}$} &  
\multicolumn{1}{c}{\TSYS $^{\rm b)}$} & 
\multicolumn{1}{c}{$\theta_{\rm B}$} & \multicolumn{1}{c}{$\eta_{\rm MB}$} & 
\multicolumn{3}{c}{Spectrometer} & \multicolumn{1}{c}{Date} \\
     & \multicolumn{1}{c}{} & & & \multicolumn{1}{c}{} & 
\multicolumn{1}{c}{} & & \multicolumn{1}{c}{Type$^{\rm c)}$} 
& \multicolumn{1}{c}{Bandwidth} & \multicolumn{1}{c}{Resolution} & \\
      & \multicolumn{1}{c}{GHz} & & & \multicolumn{1}{c}{K} & 
\multicolumn{1}{c}{$''$} & & \multicolumn{1}{c}{} 
& \multicolumn{1}{c}{MHz} & \multicolumn{1}{c}{kHz} & \\
\hline
C$^{18}$O($1\to 0$)  & 109.782160 & SEST & Schottky SSB & 400 & 46 & 0.72 & 
                  AOS         & 86   &  80          & 9/90,7/92 \\
C$^{18}$O($2\to 1$)  & 219.560319 & SEST & SIS SSB      & 700--1100 & 24 & 0.55 &
                  AOS         & 995  &  1400        & 7\&9/92 \\
                 &        & JCMT$^{\rm d)}$ & SIS DSB      & 600 & 20 & 0.69 &
                  DAS          & 250  &  189         & 4/96 \\
C$^{18}$O($3\to 2$)  & 329.350500 & JCMT & SIS DSB      & 1600--1800 & 14 & 0.58 &
                  DAS          & 500  &  378         & 4/96  \\
$^{28}$SiO($2\to 1$) & 86.846998  & SEST & Schottky SSB & 350--400 & 57 & 0.75 &
                  AOS         & 86   &  80          & 10/94 \\
$^{28}$SiO($5\to 4$) & 217.104935 & HHT  & SIS DSB      & 700 & 33 & 0.83 &
                  AOS         & 990  &  1000         & 4/97  \\
                 &     & JCMT$^{\rm e)}$ & SIS DSB      & 600 & 20 & 0.69 &
                  DAS          & 250  & 189          & 4/96  \\
$^{29}$SiO($2\to 1$) & 85.759132  & SEST & SIS SSB      & 140 & 57 & 0.75 &
                  AOS         & 86/995  & 80/1400           & 3/96  \\
$^{30}$SiO($2\to 1$) & 84.746036  & SEST & SIS SSB      & 140 & 57 & 0.75 &
                  AOS         & 86/995   & 80/1400           & 3/96 \\
\hline \\
\end{tabular} \\
a): SSB: Single sideband receiver, with a sideband rejection of $\sim 20$\,dB
; DSB: Double sideband receiver \\
b): on a \TMB\ scale \\
c): AOS: Acousto-Optical Spectrometer; DAS: Dutch Autocorrelation 
Spectrometer \\
d): Only three sources were observed for cross-telescope calibration with 
SEST \\
e): Only two sources were observed for cross-telescope calibration with the
HHT 
\end{flushleft}
\end{table*}

Typically, the telescope pointing was checked every three hours. At SEST,
two strong SiO masers close to the Galactic center, AH\,Sco and VX\,Sgr,
were measured. At the JCMT, the pointing was done using five-point continuum
maps on NGC\,6334\,I. At the HHT, continuum cross scans on Sgr\,B2 were 
used. At SEST and JCMT, the pointing accuracy was always better than $5''$. 
At the HHT, the accuracy was always better than $10''$ and usually $5''$.

At all telescopes, calibration was done with the standard chopper wheel 
method, giving temperatures on the $T_A^*$ scale. These were converted 
to \TMB , the main beam brightness temperature (see, e.g., Downes 1989 
and Rohlfs \& Wilson 1996 for definitions), using the main beam 
efficiencies, $\eta_{\rm MB}$, listed in Table\,\ref{log}.

To account for structure in our extended sources, we mapped the
beam used to measure the \CEIO ($1\to 0$) line in the ($2\to 1$) and 
($3\to 2$) transitions. Due to time restrictions, this was not possible 
for the SiO($5\to 4$) line observed at the HHT. From maps obtained at 
the JCMT for two sources in SiO($5\to 4$), the intensity in a 
33$''$ beam is higher by 8\% and 15\% than in 46$''$ and 57$''$ 
beams, respectively. This is caused by the structure of the 
individual sources, and the numbers are, therefore, uncertain.
We will, however, use these as correction factors when comparing 
HHT SiO($5\to 4$) data to measurements obtained with larger beamsizes.

To adjust the calibration scales of the three telescopes, 
we have observed a number of sources in the 1.3\,mm 
band which is accessible to all instruments (see Table\,\ref{log}). 
For \CEIO ($2\to 1$), the integrated intensities 
at the 15\,m-telescopes SEST and JCMT agree to better than 10\%, i.e.\ 
to within the uncertainty of the calibration of the individual measurement. 
The SiO($5\to 4$) JCMT maps, convolved to the larger size 
of the HHT beam, also give intensities that are consistent to within 10\%
with the HHT data. Thus, the \TMB -scales are in excellent agreement, 
and we are confident that our complete data set has been placed on a 
compatible intensity scale. 

We carried out all observations employing position switching.
This was necessary given the large angular extent of the sources, especially 
in \CEIO . Offset positions were typically chosen $15'$ (SiO) or $30'$ 
(\CEIO ) away from the source, perpendicular to the Galactic plane. This
observing mode causes baselines that are not always perfectly flat. Thus,
polynomial baselines of order $\leq 3$ were subtracted from the data.
Baselines were determined over a region of $\pm 150 - 200$\,\kms\ from 
the line center. For the JCMT data, which were observed in exceptionally 
good weather ($\tau_{225} < 0.05$ throughout), linear baselines were 
usually sufficient. 

For the final analysis, all spectra were smoothed to a velocity resolution
of $\sim 2$\,\kms . 

\section{Results}

The 33 sources we study in detail are selected as distinct peaks in the
CS survey of Bally et al.\ (1987). They are, therefore, cores that should 
have large amounts of dense gas, which are embedded in a smooth, lower
intensity molecular intercloud medium. Since we wish to derive the 
properties of typical clouds in the Galactic center environment, 
the non-typical molecular peaks associated with Sgr\,A and Sgr\,B2 
were not included in the sample. 

\begin{figure*}
\psfig{file=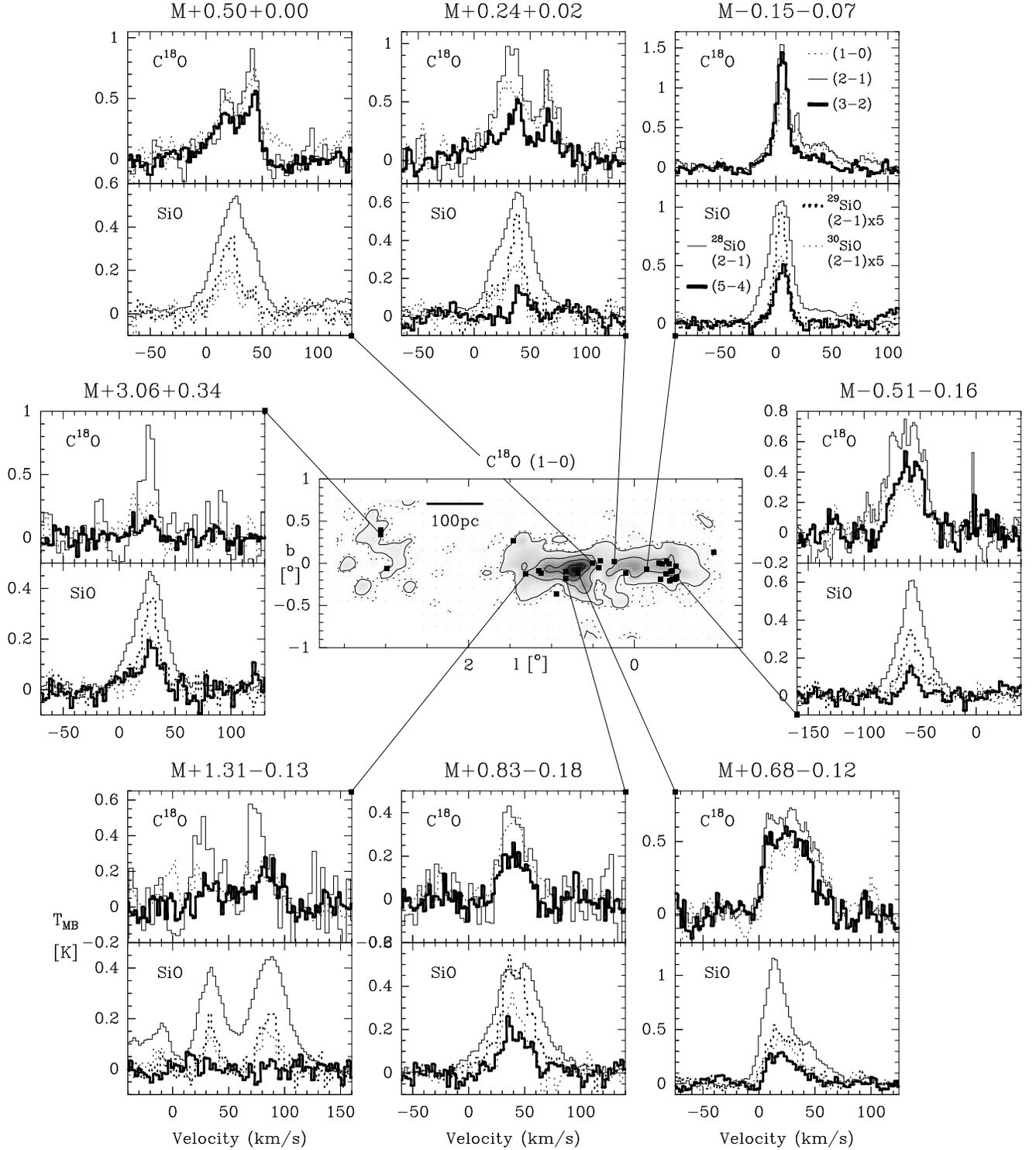,bbllx=0pt,bblly=55pt,bburx=490pt,bbury=700pt,height=21.0cm}
\caption{\label{spectra} Sample spectra of eight sources with very complete 
observations. In the top panels, the \CEIO\ lines are
displayed: The ($1\to 0$) transition is drawn as a dotted line, the ($2\to 1$) 
transition is drawn as a thin solid line and the ($3\to 2$) 
transition is displayed as a thick solid line. The bottom panels 
show the SiO spectra. $^{28}$SiO($2\to 1$) is denoted as a
thin solid line, $^{28}$SiO($5\to 4$) is given by a thick solid 
line; the $^{29}$SiO($2\to 1$) spectrum is drawn as a thick dotted line 
and the $^{30}$SiO($2\to 1$) is shown as a thin
dotted line. The intensities of both rare isotopomers were multiplied by 5
to allow clearer comparisons.  Black squares overlayed on the map of the
overall distribution of the $J=1\to 0$ line of \CEIO\ (taken from Paper\,I)
mark all 33 sources we observed except Clump\,1 at $l \sim -5^{\circ}$. }
\end{figure*}

The ($1\to 0$) and ($2\to 1$) transitions of \CEIO\ and the ($2\to 1$) 
transition of $^{28}$SiO were observed toward all sources of our 
sample. Both the \CEIO\ and (more notably) the SiO lines were 
easily seen in all sources. We observed the 10 strongest \CEIO\ 
sources in the ($3\to 2$) submillimeter transition of \CEIO , 
again detecting all. The rare isotopomers $^{29}$SiO and $^{30}$SiO 
were observed and detected toward 12 clouds showing strong emission 
in $^{28}$SiO. Of 8 sources measured in the $^{28}$SiO($5\to 4$)
line, 7 showed emission.  In Fig.\,\ref{spectra} and Table\,\ref{rat}, 
we give an overview of our results. Sample spectra for eight sources 
are displayed in Fig.\,\ref{spectra}. We mark the position  of all 
sources observed on a survey map of the ($1\to 0$) transition of 
\CEIO\  (Papers I and II). Galactic coordinates are used for 
source names.  For the corresponding equatorial coordinates, refer 
to H\"uttemeister et al.\ (1993b). Most of the clouds we have studied are located within the molecular bulge region, while three lie within 
the `Clump\,2' complex (Bania 1977, Stark \& Bania 1986) at $l 
\sim 3^{\circ}$. `Clump\,1' is located $\sim 5^{\circ}$ south of 
the Galactic center. 

Integrated line intensities for the \CEIO ($2\to 1$) and 
$^{28}$SiO($2\to 1$) transitions and line intensity ratios derived 
from integrated intensities are presented in Table\,\ref{rat}. 
The errors are determined from the (usually small) formal error of a 
Gaussian fit to the lines or, in case of non-Gaussian lineshapes, from 
the rms noise in the spectrum. In addition, a calibration uncertainty of 
10\% for data taken with the same telescope  or 20\% for data obtained 
with different instruments was assumed. For Gaussian lines, the formal 
error of the fit agrees closely with the error obtained from the rms 
noise in the spectrum. 

\begin{table*}
\caption{ \label{rat} Integrated intensities of the \CEIO ($2\to 1$) and
SiO($2\to 1$) transitions and line intensity ratios of \CEIO\ and SiO.} 
\begin{tabular}{l|rrll|rrlll}
Source        
              & $v_{\rm LSR}^{\rm C^{18}O}$ 
              & $I_{(2-1)}^{\rm C^{18}O}$ 
              & ${\cal R}_{2/1}^{\rm C^{18}O}$ 
              & ${\cal R}_{3/2}^{\rm C^{18}O}$
              & $v_{\rm LSR}^{\rm SiO}$
              & $I_{(2-1)}^{\rm ^{28}SiO}$ 
              & ${\cal R}_{5/2}^{\rm ^{28}SiO}$  
              & $r_{28/29}^{\rm SiO}$ 
              & $r_{29/30}^{\rm SiO}$ \\
              & & & & 10$^{15}$ 
              & & & & & 10$^{13}$  \\
              & km/s & K\,km/s & & 
              & km/s & K\,km/s & & &  \\
        \hline
Clump\,1      &     97 & 4.5(0.2) & 1.46(0.27)  & & 97 & 0.6(0.2) & & \\
M--0.96+0.13  &    135 & 5.1(0.5) & 1.31(0.24)$^{\rm a)}$  &  
              &    135 & 4.6(0.3) & & &  \\
M--0.50--0.03 &  --100 & 15.3(0.4) & 1.44(0.16)  & 0.49(0.11)  
              &   --98 & 4.0(0.3) & & &  \\
              &    --2 & 1.6(0.2) & 1.12(0.31)$^{\rm c)}$ & 0.38(0.11)$^{\rm c)}$ 
              &    --2 & $\leq 0.1$ & & &  \\
M--0.42--0.01 &   --84 & 10.5(0.4) & 1.05(0.12) &  
              &   --80 & 3.4(0.2) & &  \\
              &    --4 & 2.2(0.3) & 1.35(0.34)  &  
              &    --4 & $\leq 0.2$ & & &  \\
M--0.39+0.02  &   --80 & 13.8(1.2) & 1.89(0.27)  &  
              &   --79 & 3.3(0.2) & & &  \\
M--0.33--0.01 &   --47 & 12.8(0.5) & 1.19(0.18)  &  
              &   --51 & 4.4(0.2) & & &  \\
M--0.45--0.09 &   --61 & 4.0(0.4) & 0.77(0.25)$^{\rm a)}$ &  
              &   --61 & $\leq 0.3$ & -- & -- &  \\
              &   --10 & 11.4(0.5) & 2.09(0.39) &  
              &   --12 & 16.7(1.5) & 0.23(0.05) & 15.9(3.1) &  \\
M--0.30+0.00  &   --33 & 12.1(0.7) & 1.23(0.25)$^{\rm b)}$  & 
              &   --38 & 5.0(0.3) & & &  \\
M--0.44--0.10 &   --15 & 6.1(0.6) & 0.60(0.31)$^{\rm a),b)}$  &  
              &   --20 & 19.8(2.0) & & 12.3(1.8) & 1.7(0.3)  \\
M--0.51--0.16 &   --61 & 27.2(0.8) & 2.67(0.14)  & 0.51(0.08)  
              &   --57 & 30.6(0.5) & 0.09(0.03) & 11.0(1.4) & 1.6(0.3)  \\
              &    --2 & 1.7(0.2) & 1.33(0.75)$^{\rm c)}$  
                    & 0.41(0.14)$^{\rm c)}$ 
              &    --2 & $\leq 0.1$ & -- & -- & --  \\
M--0.50--0.18 &   --57 & 14.3(0.7) & 2.40(0.67)  &  
              &   --55 & 9.3(0.4) & & &  \\
M--0.48--0.19 &   --50 & 6.9(0.4) & 1.47(0.32)$^{\rm a)}$  &  
              &   --53 & 4.0(0.2) & & &  \\
M--0.38--0.13 &   --33 & 10.4(0.7) & 1.11(0.26)  & 0.57(0.12)  
              &   --32 & 1.7(0.2) & & &   \\
M--0.45--0.19 &     21 & 4.5(0.2) & 1.10(0.19)$^{\rm c)}$  &  
              &     21 & $\leq 0.2$ & & & \\
              &     58 & 3.9(0.3) & 0.50(0.31)$^{\rm a)}$  &  
              &     58 & 5.8(0.2) & & &  \\
M--0.43--0.21 &   --34 & 22.4(0.9) & 2.07(0.27)  &  
              &   --27 & 6.5(0.3) & & &  \\
              &     19 & 3.5(0.3) & 1.13(0.21)$^{\rm c)}$ &  
              &     19 & $\leq 0.2$ & & &  \\
              &     53 & 2.7(0.6) & 0.71(0.22)$^{\rm a)}$ &  
              &     54 & 5.7(0.4) & & &  \\
M--0.15--0.07 &      5 & 37.5 (0.4) & 1.40(0.15)$^{\rm b)}$ 
                       & 0.75(0.14)$^{\rm b)}$ 
              &      4 & 31.5(1.1) & 0.19(0.04)$^{\rm b)}$ & 11.3(1.3) 
                       & 1.6(0.2) \\
M--0.32--0.19 &   --32 & 7.3(0.7) & 1.22(0.28)$^{\rm a)}$  & & & & & \\
              &     25 & $\leq$3.0 & & & 25 & 2.9(0.2) & & &  \\
M+0.24+0.02   &     35 & 41.7(3.7) & 1.06(0.16)$^{\rm b)}$ 
                       & 0.51(0.11)$^{\rm b)}$ 
              &     36 & 24.4(0.7) & 0.07(0.02) & 11.4(1.3) &  1.6(0.3) \\
M+0.10--0.12  &     22 & 22.1(0.9) & 1.46(0.21)  &  
              &     29 & 11.5(1.2) & & &  \\
M+0.43--0.05  &    --2 & 4.5(0.3) & 0.95(0.28)$^{\rm a)}$  &  
              &     12 & 11.0(0.8) & & 13.3(2.2) &  \\
              &     38 & 4.9(0.3) & 0.69(0.15)$^{\rm a)}$  & & & & & \\
M+0.41+0.03   &     30 & 10.5(0.4) & 0.64(0.10)$^{\rm b)}$  & 
              &     21 & 10.4(1.1) & & &  \\
M+0.50+0.00   &     30 & 22.9(2.4) & 1.06(0.21)$^{\rm b)}$  
                       & 0.76(0.15)$^{\rm b}$  
              &     26 & 20.9(0.3) & & 13.1(2.1) & 1.7(0.3)   \\
M+0.68--0.12  &     30 & 35.4(0.3) & 1.48(0.17)$^{\rm b)}$  
                       & 0.70(0.14)$^{\rm b)}$ 
              &     20 & 39.2(2.3) & 0.18(0.05)$^{\rm b)}$ & 11.9(1.4) 
                       & 1.2(0.2) \\
M+0.83--0.10  &     24 & 13.4(0.8) & 0.81(0.18)$^{\rm a),b)}$  &  
              &     24 & 10.4(0.4) & & &  \\
              &     99 & 8.7(0.5) & 1.66(0.29) &  
              &     96 & 2.5(0.2) & & &  \\
M+1.46+0.27   &     80 & 6.6(0.4) & 0.95(0.32)$^{\rm a)}$ &  
              &     81 & 20.4(0.3) & & 13.8(1.6))$^{\rm b)}$ & 1.5(0.3)  \\
M+0.83--0.18  &     41 & 10.6(1.0) & 1.03(0.13)  & 0.53(0.12) 
              &     43 & 26.0(0.4) & 0.19(0.04) & \enspace7.4(0.8) & 2.0(0.2) \\
M+1.15--0.09  &   --18 & 1.9(0.2) & 1.05(0.20)$^{\rm c)}$ &  
              &   --18 & $\leq 0.2$ &  &  &  \\
              &     60 & 7.8(1.2) & 0.68(0.23)$^{\rm a),b)}$ &  
              &     67 & 21.0(1.1) & & &  \\
M+1.13--0.12  &   --15 & 12.9(0.9) & 1.48(0.18) &  
              &   --15 & 6.4(0.7) & & &   \\
              &     80 & 1.8(0.3) & 0.47(0.22)$^{\rm a)}$ &  
              &     70 & 17.5(1.2) & & &  \\
M+1.31--0.13  &     30 & 11.6(0.6) & 1.95(0.39)$^{\rm a)}$ 
                       & 0.25(0.13)$^{\rm a)}$
              &     35 & 11.5(0.5) & $\leq 0.04$$^{\rm d)}$  & 20.9(2.3) 
                       & 1.5(0.3)  \\
              &     80 & 7.8(0.6) & 1.58(0.35)$^{\rm a)}$ 
                       & 0.78(0.26)$^{\rm a)}$ 
              &     87 & 16.8(0.5) & $\leq 0.04$$^{\rm d)}$ & 18.0(2.7) 
                       & 1.4(0.2)   \\
M+0.94--0.36  &   --49 & 18.5(1.3) & 2.05(0.33)$^{\rm b)}$ &  
              &   --41 & 3.4(0.5) & & &  \\
M+3.06+0.40   &     12 & 1.6(0.1) & 0.82(0.15)$^{\rm c)}$ & 
              &     12 & $\leq 0.1$ & &  \\
              &     22 & 2.5(0.3) & 0.61(0.16)$^{\rm a)}$ &  
              &     30 & 9.3(2.1) & & &   \\
M+3.06+0.34   &     27 & 12.1(3.6) & 1.49(0.48)  & 0.24(0.09)  
              &     27 & 15.2(0.3) & 0.16(0.06) & 10.1(1.2)  & 1.9(0.4)  \\
M+2.99-0.06   &   --13 & 2.6(0.3) & 1.21(0.27)  &  
              &   --12 & 8.5(0.3) & & &  \\
\hline \\
\end{tabular} \\
Abbreviations used throughout the paper: 
\RZWCO\,$=\,I_{(2-1)}^{\rm C^{18}O}/I_{(1-0)}^{\rm C^{18}O}$,
\RDRCO\,$=\,I_{(3-2)}^{\rm C^{18}O}/I_{(2-1)}^{\rm C^{18}O}$,
\RMSIO\,$=\,I_{(5-4)}^{\rm ^{28}SiO}/I_{(2-1)}^{\rm ^{28}SiO}$,
\RISOSIO\,$=\,I_{(2-1)}^{\rm ^{28}SiO}/I_{(2-1)}^{\rm ^{29}SiO}$,
 $r_{29/30}^{\rm SiO}\,=\,I_{(2-1)}^{\rm ^{29}SiO}/I_{(2-1)}^{\rm 
^{30}SiO}$ \\
a): The result is uncertain, due to either baseline uncertainties or low 
    signal-to-noise ratios \\
b): Non-gaussian lineshape or several blended velocity components;
    ratios have been calculated by adding channels with emission \\
c): very narrow line, possibly from local ISM \\
d): 3$\sigma$  \\
\end{table*}

For a given species, the center velocities and lineshapes of the 
different transitions always agree, to within the noise. 
Between SiO and \CEIO , however, there can be significant 
differences. The sources M+0.50+0.00 and M+0.24+0.02 are good examples: 
In both cases, the \CEIO\ lines show two distinct peaks, while 
the SiO transitions are single-peaked. In M+0.50+0.00, the central 
velocity of the SiO is close to the weaker \CEIO\ line; in M+0.24+0.02 
SiO and the stronger \CEIO\ peak agree.
It is also noteworthy that narrow \CEIO\ lines close
to a \vlsr\ of 0\,\kms , likely of local origin, never have a 
counterpart in SiO. This is illustrated by the source M--0.51--0.16 in 
Fig.\,\ref{spectra} and demonstrates the unusual nature of the Galactic 
center sources as compared to Galactic disk clouds. 

Since the lines of $^{29}$SiO and $^{30}$SiO are very likely optically
thin, we can directly check whether the ratio of $^{29}$Si/$^{30}$Si 
in the Galactic center region agrees with the terrestrial value. 
We find a line intensity ratio of $1.6 \pm 0.2$ in our sample, in 
excellent agreement with the terrestrial isotope ratio of 1.5. 
This confirms that this ratio does not depend on the galactocentric distance 
(Wilson \& Rood 1994, Penzias 1981), which is an expected result if 
both isotopes are synthezised in the s-process in stars 
of the same type. 

For $^{28}$SiO/$^{29}$SiO, we 
take the ratio to be the terrestrial value of 20, as suggested by 
Penzias (1981). Since the $^{28}$SiO($2\to 1$) transition is 
optically thick, we cannot check this assumption, but it is supported
by the largest line ratio \RISOSIO\ (toward the exceptional source 
M+1.31--0.13, see Section 4.3), which is indeed close to 20.

\section{The physical conditions of the gas}

\subsection{Determinination of densities and temperatures}

A Large Velocity Gradient (hereafter LVG) model was used to estimate 
physical conditions of the gas, namely density (\numd ), kinetic 
temperature (\TKIN ) and -- in the case  of SiO -- molecular 
abundance  ($X$(SiO)) (see Scoville \& Solomon 1974 and De Jong et al.\ 
1975 for the general properties). We used a modified version of the code 
of Henkel et al.\ (1980). The basic LVG assumption is that a of 
systematic velocity gradient large compared to random motions and 
monotonic. This allows us to treat the molecular excitation as a local 
problem. This is certainly an idealisation for Galactic center clouds. 
However, an application of an LVG code requires no detailed knowledge of
the velocity field. In the case of CO lines, H$_2$ densities 
determined under the assumption of LVG and microturbulence (the opposite
extreme) do not differ by more than a factor of three (White 1977). 

Our analysis is also based on the assumption that, for a given 
molecule, all transitions observed arise in the  same volume. 
Since the excitation depends on critical density and optical depth, 
and thus is not identical for different transitions or different 
isotopomers, this assumption may not be strictly correct, 
especially if the medium is very clumpy (see, e.g., the discussion 
in Oka et al.\ 1997 for $^{12}$CO). For \CEIO , this is not a 
critical problem since the LVG calculations show that all 
transitions measured by us are optically thin, or, in 
the low temperature/high density scenario described 
below, reach, at most, $\tau \approx 1$. Therefore, \CEIO\ line 
intensity ratios are not strongly affected by the  possibly different 
locations of cloud `photospheres' from which the bulk of the photons 
are emitted. All transitions of  \CEIO\  should trace all molecular gas 
at densities $\geq 10^3$\,\percc, as long as \CEIO\ is not selectively
dissociated by UV radiation.  Away from the star forming regions Sgr\,A 
and Sgr\,B2, the UV  radiation field in the Galactic center region is not 
likely to be strong (e.g.\ Nagakawa et al.\ 1995).  In addition, 
because we are considering  density peaks of the molecular gas (and dust), 
the destruction of C$^{18}$O should not be a problem; from the 
H$_2$ column densities derived below, dust extinction is likely to provide 
sufficient shielding. 

\begin{figure}
\psfig{file=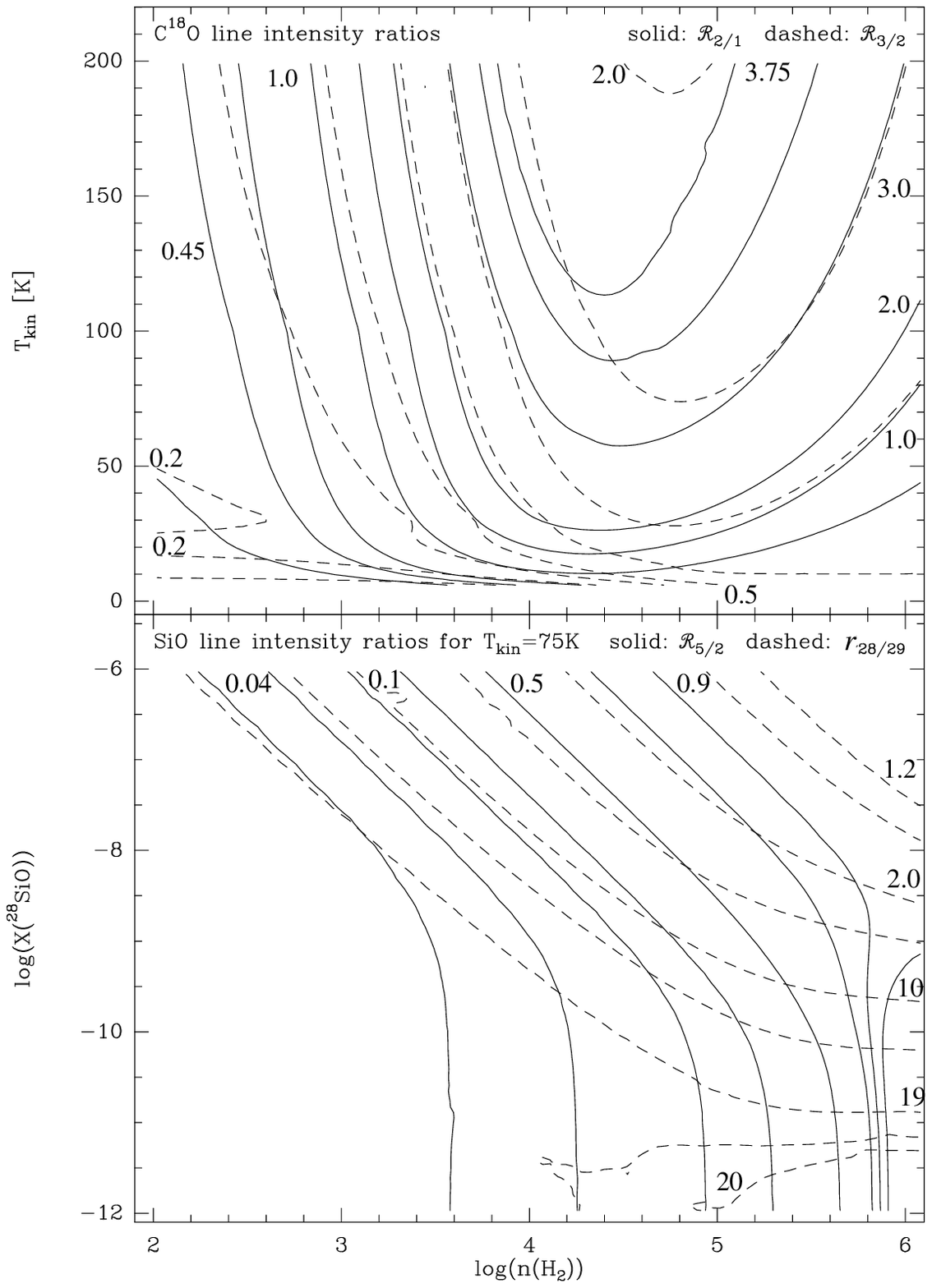,bbllx=0pt,bblly=40pt,bburx=290pt,bbury=500pt,height=10.5cm}
\caption{ \label{lvg1} Line intensity ratios from our LVG model calculations
for \CEIO\ (upper panel) and $^{28}$SiO and $^{29}$SiO (lower panel). 
The contours are at: 
\protect\\
\RZWCO (solid): 0.3, 0.45, 0.6, 1.0, 1.5, 2.0, 3.0, 3.5, 3.75 
\protect\\
\RDRCO (dashed): 0.1, 0.2, 0.35, 0.5, 0.75, 1.0, 1.5, 2.0 
\protect\\
\RMSIO (solid): 0.04, 0.05, 0.1, 0.2, 0.5, 0.8, 0.9, 1.0 
\protect\\
\RISOSIO (dashed): 1.2, 1.5, 2.0, 5.0, 10.0, 15.0, 19.0, 19.75, 20.0}
\end{figure}

In our model calculations, the population $n(J)$, excitation temperature 
\TEX , brightness temperature \TB , and optical depth $\tau$, are
determined for each rotational level up to a maximum $J= 20$ for 
\CEIO\ and $J= 13$ for SiO. Collision rates are taken from Green \&
Chapman (1978). We have chosen a constant velocity 
gradient of 5\,\kms\,pc$^{-1}$, a value estimated from the velocity 
extent and diameter of Galactic center clouds as mapped by Bally et al.\ 
(1987) or Lindquist et al.\ (1995). The background continuum temperature 
was assumed to be 2.7\,K.

\subsubsection{C$^{18}$O}

We assume that the abundance of \CEIO\ relative to H$_2$ is  $4
\cdot 10^{-7}$. This ratio is based on the standard Galactic center
$^{16}$O/$^{18}$O isotopic ratio of 250 (Wilson \& Matteucci 1992, 
Wilson \& Rood 1994) and a  CO/H$_2$ ratio of 10$^{-4}$ (e.g.\ Frerking
et al.\ 1982, Duvert et al.\ 1986 (dark clouds), Blake et al.\ 1987 
(OMC-1)). According to the calculations of Farquhar et al.\ (1994), this
ratio is stable against a possibly enhanced cosmic ray flux close to
the Galactic center. For \CEIO , we then calculate a grid by varying 
the H$_2$ density, \numd , from $10^{2}$\,\percc\ to $10^{6}$\,\percc\ and  
the kinetic temperature, \TKIN , from 5 to 200\,K. 

We found the range of \numd\ and \TKIN\  
by comparing the observed line intensity ratios (Table\,\ref{rat})
to the ratios predicted by the model (Fig.\,\ref{lvg1}). It is 
important to use line intensity ratios, not just intensities, since
intensities are affected by an a priori unknown beam 
filling factor, $f_{\rm b}$. For the intensity ratios, we took 
$f_{\rm b}$ to be identical for all transitions of a given molecule.

The integrated intensity ratio of the \CEIO ($2\to 1$) to ($1\to 0$) 
transition, \RZWCO , from all sources is $1.3 \pm 0.5$, with extreme \
values ranging from 0.5 to $\sim 2$. The ratio of the ($3\to 2$) to the 
($2\to 1$) intensity, \RDRCO , is $0.5 \pm 0.2$. Averaged over all 
sources, the LVG model fits yield \TEX\,$= 11( \pm 4)$\,K for the \CEIO 
($2\to 1$) transition, that is, \CEIO\ is subthermally excited. The 
beam averaged peak optical depth $\tau_{21}$ is always $< 0.2$. The 
measured peak temperatures \TMB\ indicate a beam filling factor 
$f_{\rm b}^{\rm C^{18}O}$ of $\sim 1$ for \numd\,$=(2 - 
4)\,10^3$\,\percc .  $f_{\rm b}^{\rm C^{18}O}$ is lower  
if \numd\ is higher. From the structure seen in our maps of the \CEIO 
(1$\to$0) beam in higher transitions, we estimate $f_{\rm b}^{\rm C^{18}O}$, 
averaged over all gas components, to be $\sim 0.5$. These maps, however,
have a low signal-to-noise ratio. 

From the model calculations (Fig.\,\ref{lvg1}, upper panel), maximum values 
up to 4 in \RZWCO\ and up to 2 in \RDRCO\ are expected for hot (\TKIN $> 
100$\,K) and dense (\numd\,$\sim 10^{4} - 10^{5}$\,\percc) gas. Such 
high ratios are reached in none of our sources. Possible combinations of
\numd\ and \TKIN\ are {\em either} a high \TKIN\ of $\geq 100$\,K 
and \numd\ of (1 -- 4)\,10$^3$\,\percc\ {\em or} a lower 
\TKIN\ ($\leq 20$\,K -- 30\,K) at a higher \numd\,$\geq 10^4$\,\percc. 
We have carried out this analysis individually for all
clouds. For all sources our \CEIO\ data {\em unambiguously} require gas with 
densities of $\geq 10^4$\,\percc\ in these Galactic center clouds 
to be {\em cool}.

From ammonia measurements (H\"uttemeister et al.\ 1993b) it is known
that both hot and cool gas is present within the area covered by our 
beam. Thus, it is likely that both scenarios possible from the \CEIO\
data are realized. A continuum of temperatures and densities is compatible 
with our data. Then \TKIN\ rises steadily as \numd\ decreases, when the line
of sight samples different parts of the cloud. Density peaks (indicated 
by strong CS lines) in Galactic center region GMCs not presently 
undergoing massive star formation can be regarded as {\em `cool dense cores'} 
in contrast to the hot dense cores generally encountered within GMCs in 
the disk. 
  
\subsubsection{SiO}
It has been shown that the fractional abundance of SiO, $X$(SiO), can 
change by more than six orders of magnitude between quiescent cold 
material and hot, shocked dense gas (Ziurys et al.\ 1989,
Mart\'{\i}n-Pintado et al.\ 1992). Since SiO is readily detected in 
all our sources, $X$(SiO) {\em cannot} be low. However, we 
cannot assume any specific value. Therefore, we have used $X$(SiO)
as a free parameter in the models, varying it from $10^{-12}$ to $10^{-6}$. 
Grids have been calculated as a function of \numd\ and $X$(SiO) for 
\TKIN $=$ 25\,K, 50\,K, 75\,K, 100\,K, 150\,K.

The average line intensity ratio between the SiO\,$(5\to 4$) and ($2\to 1$) 
transition is \RMSIO\,$=\,0.16 \pm 0.05$, for those sources where we 
detected SiO($5\to 4$). M+1.31--0.13 is exceptional and will be 
discussed further in Section\,4.3. Since the signal-to-noise (S/N) 
ratio in the ($5\to 4$) line is not always high and the ($2\to 1$) SEST 
beam was not mapped in this transition, it is not clear whether  
source to source differences (apart from M+1.31--013) are significant. 
For the line ratios between the ($2\to 1$) transitions of 
$^{28}$SiO and $^{29}$SiO (excluding M+1.31--0.13), we find a 
mean ratio of \RISOSIO$\,=\,12.2\pm 2.2$ with extreme values
ranging from 7.4 to 15.9. Since these data, even in the $^{29}$SiO 
transition, have very good S/N ratios and were measured with the 
same telescope and beamsize, there is no doubt that the 
source-to-source differences are significant. 

Fig.\,\ref{lvg1} (lower panel) shows the line ratios \RMSIO\ 
and \RISOSIO\ for \TKIN = 75\,K from our LVG calculations. Lowering 
\TKIN\ to 25\,K requires a slight increase 
of the H$_2$ densities for a given line ratio, by a factor of $\sim 1.25$. 
Raising \TKIN\ to 150\,K corresponds to slightly lower densities. 
In general, the effects of changing \TKIN\ on the line ratios are 
negligible: SiO line ratios are almost insensitive to cloud temperature.  

To derive \TEX\ and \numd , we have analysed all sources individually. 
Excluding M+1.31--0.13, we find \TEX\ ranging from 5\,K to 10\,K. 
\numd\ changes with the assumed SiO abundance: A higher $X$(SiO) 
requires a lower H$_2$ density (Fig.\,\ref{lvg1}). Curves of
constant \RMSIO\ and \RISOSIO\ run parallel for a range of densities and 
SiO abundances. Typical ranges are  $X$(SiO)\,$\approx 10^{-6}$ for
\numd\,$\approx (1 - 2.5)\,10^3$\,\percc\ to $X$(SiO)\,$\approx 10^{-9}$
for \numd\,$\approx 4\,10^4 - 10^5$\,\percc . Lower SiO abundances 
or higher densities are not possible, since then the curves diverge. 

Plotting the measured \TMB 's for the $^{28}$SiO($2\to 1$) 
transition in the LVG plots (e.g.\ Fig.\,\ref{lvg}, lower left panel),
we find that a considerably higher \TB\ is needed to fit the observed 
line ratios. Thus we estimate that the beam filling factor 
$f_{\rm b}^{\rm SiO}$ (\TMB\,=\,$f_{\rm b}^{\rm SiO}$\,\TB ) is
$\sim 0.2$ for the `typical' sources. 

This analysis is biased toward more intense SiO lines, since only 
those were observed in transitions other than $^{28}$SiO$(2\to 1)$. 
The strength in SiO toward these positions may be either due to a 
large amount of molecular gas, or to an abundance of SiO that is 
above average. This question will be addressed in the next section. 

\subsection{A joint view of C$^{18}$O and SiO}

\subsubsection{Column densities and SiO abundances} 

Further insight in the structure of the clouds is gained
by combining what can be learned from \CEIO\ and SiO. 

We calculate the total (beam averaged) column densities for \CEIO\ and 
$^{28}$SiO, applying the same procedure to both molecules: First, we use
the observed \TMB\ and the best fit \TEX\ from the LVG model 
for the ($2\to 1$) transitions (the lines with the best S/N ratio) to
calculate the (beam averaged) optical depth $\bar \tau_0$. Taking 
$ \int\,\bar \tau(v)\,{\rm d}v\,= 1.06\,\bar \tau_0\, \Delta v_{1/2}$, we then 
derive the column density in the $J=1$ level, $\bar N_1$. The total beam 
averaged column density $\bar N$ is determined by dividing $\bar N_1$
by the fraction of the population residing in the $J=1$\ level, derived
from the LVG model. (See Rohlfs \& Wilson (1996) for a collection of the 
(standard) formulae we used.) For non-gaussian lines, the second moment of
the line has been used as a measure of the line width instead of 
$\Delta v_{1/2}$. This has been combined with a \TMB\ which reproduces 
the observed integrated intensity. This is correct as long as \TEX\ 
does not change across the line profile. Since the line shapes in 
all transitions observed for a given molecule agree closely, this is
a reasonable assumption. 

For the sources where we observed only the SiO($2\to 1$) line, we have
used average values for \TEX\ and the fraction of the population in 
the $J = 1$ level. These are 7\,K and 0.36, respectively.
Total beam averaged column densities are given in Table\,\ref{col}. 

\begin{table}[t!]
\caption{ \label{col} Total beam averaged column densities of 
\CEIO\ ($\bar N^{\rm C^{18}O}$) and SiO ($\bar N^{\rm SiO}$) and 
beam averaged abundances of SiO, $X$(SiO). $\bar N^{\rm H_2}$ follows
from $\bar N^{\rm C^{18}O}$, assuming $X$(\CEIO ) to
be $4 \cdot 10^{-7}$. Errors are given in parentheses.}
\begin{tabular}{l|rr|rrr}
Source        & $v_{\rm LSR}^{\rm C^{18}O}$ 
              & $\bar N^{\rm C^{18}O}$  
              & $v_{\rm LSR}^{\rm SiO}$
              & $\bar N^{\rm SiO}$ 
              & $X^{\rm SiO}$      \\
              & &  10$^{15}$ 
              & & 10$^{13}$ 
              & 10$^{-9}$    \\
              & km/s & cm$^{-2}$ 
              & km/s & cm$^{-2}$ &  \\
        \hline
Clump\,1      &     97 &  2.6(0.9) & 97  & $\sim 0.1$ & 0.1\\
M--0.96+0.13  &    135 &  2.9(1.1) 
              &    135 &  $\sim 0.9$ & 1.7 \\
M--0.50--0.03 &  --100 &  8.9(1.4) 
              &  --98  &  $\sim 0.8$ & 0.4 \\
              &    --2 &  1.1(0.2) 
              &   --2  & $< 0.02$ & $<0.07$\\
M--0.42--0.01 &   --84 &  7.5(0.8) 
              &   --80 & $\sim 0.6$ & 0.3 \\
              &    --4 & 1.2(0.1) 
              &    --4 & $< 0.1$ & $<0.03$ \\
M--0.39+0.02  &   --80 & 6.6(0.6) 
              &   --79 & $\sim 0.6$ & 0.4 \\
M--0.33--0.01 &   --47 & 7.7(2.5) 
              &   --51 & $\sim 0.8$ & 0.4  \\
M--0.45--0.09 &   --61 & 3.6(0.9) 
              &   --61 & $< 0.09$ & $<0.1$ \\
              &   --10 & 5.5(0.5) 
              &   --12 & 3.2(0.4) & 2.3 \\
M--0.30+0.00  &   --33 & 7.2(1.5) 
              &   --38 & $\sim 1.0$ & 0.6 \\
M--0.44--0.10 &   --15 & 7.4(2.9) 
              &   --20 & 4.4(1.0) & 2.4 \\
M--0.51--0.16 &   --61 & 13.1(3.5) 
              &   --57 & 4.0(1.5) & 1.2 \\
              &    --2 & 1.0(0.3) 
              &    --2 & $< 0.03$ & $<0.1$ \\
M--0.50--0.18 &   --57 & 7.6(1.5) 
              &   --55 & $\sim 1.8$  & 1.0  \\
M--0.48--0.19 &   --50 & 3.6(0.8) 
              &   --53 & $\sim 0.7$ & 0.7 \\
M--0.38--0.13 &   --33 & 6.6(0.9) 
              &   --32 & $\sim 0.3$ & 0.2 \\
M--0.45--0.19 &     21 & 3.0(0.5) 
              &     21 & $< 0.04$ & $<0.05$ \\
              &     58 & 4.4(1.0) 
              &     58 & $\sim 1.1$ & 1.0 \\
M--0.43--0.21 &   --34 & 10.4(0.4) 
              &   --27 & $\sim 1.2$ & 0.5 \\
              &     19 & 2.2(0.3) 
              &     19 & $< 0.04$ & $<0.07$ \\
              &     53 & 2.8(0.6) 
              &     54 & $\sim 1.1$ & 0.6 \\
M--0.15--0.07 &      5 & 19.1(1.0) 
              &      4 & 6.2(1.0)  & 1.3 \\
M--0.32--0.19 &   --32 & 4.3(0.7) & & &  \\
              &     25 & $\leq 1.6$ 
              &     25 & $\sim 0.6$ & $>1.5$ \\
M+0.24+0.02   &     35 & 26.1(5.0) 
              &     36 & 5.6(1.2) & 0.9 \\
M+0.10--0.12  &     22 & 11.5(1.9) 
              &     29 & $ \sim 2.2 $ & 0.8 \\
M+0.43--0.05  &    --2 & 3.4(0.8) 
              &     12 & 2.4(0.6) & 2.8 \\
              &     38 & 4.9(0.9) & & & \\
M+0.41+0.03   &     30 & 14.1(1.0) 
              &     21 & $\sim 2.0$ & 0.6 \\
M+0.50+0.00   &     30 & 13.2(2.9) 
              &     26 & 5.0(1.6)  & 1.5 \\
M+0.68--0.12  &     30 & 17.2(2.0) 
              &     20 & 7.6(0.7) & 1.8 \\
M+0.83--0.10  &     24 & 12.1(3.5) 
              &     24 & $\sim 1.9 $ & 0.6 \\
              &     99 & 4.4(0.4) 
              &     96 & $\sim 0.5$ & 0.5 \\
M+1.46+0.27   &     80 & 4.9(1.3) 
              &     81 & 4.9(1.0)  & 4.0 \\
M+0.83--0.18  &     41 & 6.8(1.6) 
              &     43 & 4.9(0.2) & 2.9 \\
M+1.15--0.09  &   --18 & 1.2(0.2) 
              &   --18 & $<0.06$ &  $<0.02$ \\
              &     60 & 8.9(1.9) 
              &     67 & $\sim 4.0$ & 1.8 \\
M+1.13--0.12  &   --15 & 7.4(0.8) 
              &   --15 & $\sim 0.5$ & 0.3  \\
              &     80 & 3.3(1.3) 
              &     70 & $\sim 3.3$ & 4.0 \\
M+1.31--0.13  &     30 & 6.9(3.5) 
              &     35 & $\sim 10 $ & 5.8 \\
              &     80 & 4.0(0.4) 
              &     87 & $\sim 10 $  & 10.0 \\
M+0.94--0.36  &   --49 & 8.8(0.3) 
              &   --41 & $\sim 0.6$ & 0.3 \\
M+3.06+0.40   &     12 & 1.5(0.3) 
              &     12 & $< 0.02$ & $<0.05$ \\
              &     22 & 3.7(1.0) 
              &     30 & $\sim 1.7$  & 1.8  \\
M+3.06+0.34   &     27 & 10.0(3.0) 
              &     27 & 2.9(0.3) & 1.2  \\
M+2.99-0.06   &   --13 & 3.3(0.7) 
              &   --12 & $\sim 1.6$  & 1.9  \\
\hline \\
\end{tabular} \\
\end{table} 

The usually optically thin, easily excited transitions of \CEIO\ are known 
to be excellent tracers of H$_2$ column density, $\bar N_{\rm H_2}$, 
over a wide variety of H$_2$ densities and kinetic temperatures. 
In particular, the integrated intensity of the $J = 2 \to 1$ transition 
can be directly related to $\bar N_{\rm H_2}$ (Genzel 1992). Our LVG 
modelling confirms that for a large range of temperatures the fraction 
of the total population in the \CEIO\ $J = 1$ level remains almost constant 
between \numd\,$=10^2$\,\percc\ and $10^4$\,\percc. 

In most cases, we find that column densities agree with 
the formula given by Mauersberger et al.\ (1992) and  Genzel (1992) 
($\bar N_{\rm H_2} = 1.12\ 10^{21} \int T_{\rm MB}^{\rm C^{18}O} 
(2\to 1)\,{\rm d}v $ for the Galactic center isotopic ratios) to 
within 10\%. Differences up to a factor of 2 are found if the 
excitation temperature is exceptionally low. 

The knowledge of $\bar N_{\rm H_2}$ and $\bar N^{\rm SiO}$ allows us
to derive a beam averaged SiO relative abundance, assuming that the 
\CEIO\ and SiO emission arises from the same gas. This is given in 
Table\,\ref{col}, and typically ranges from 0.5\,10$^{-9}$ to
5\,10$^{-9}$. Since \CEIO\ presumably traces {\em all} H$_2$ with 
\numd\,$\geq 10^3$\,\percc, the true abundance of SiO cannot be {\em lower}
than the value thus determined. It can be higher, if only part 
of the gas within the beam is abundant in SiO, which, considering the 
differences in line centers and line profiles between \CEIO\ and SiO 
(e.g.\ M+0.24+0.02, see Fig.\,\ref{spectra}), is likely.  

For those sources where the SiO($2\to 1$) transition is 
weak and which were only observed in this SiO transition, the H$_2$ column 
density obtained from \CEIO\ is not systematically lower
than for the more intense sources. On average, in these clouds $X$(SiO)
is lower by a factor of $\sim$\,3. Source-to-source variations 
are significant, even though the properties in dense gas as traced by 
CS (H\"uttemeister 1993) and total gas as traced by \CEIO\ do not differ. 
This agrees with the variation found in the survey of Mart\'{\i}n-Pintado 
et al.\ (1997). Note, however, that we find SiO with an abundance that is 
high when compared to typical disk GMCs in all sources.  

\subsubsection{Density and temperature}

We now combine the LVG models for \CEIO\ and SiO to derive a consistent 
solution for \numd\ and \TKIN . This is done in terms of a two component
model,  where part of each \CEIO\ emission line is ascribed to a cool 
component, with the remaining part arising from a hot component. 

From the NH$_3$ studies of H\"uttemeister et al.\ (1993b), we know that,
for a typical cloud, roughly 25\% of the neutral gas mass is contained in the 
hot ($> 100$\,K) component, which, from our \CEIO\ analysis, must have a low 
density. Thus, the remaining $\sim 75$\% of the total gas mass is at a 
density of $\geq 10^4$\,\percc. The hot, thin and the cool, dense component
are roughly in pressure equilibrium. Let us assume that the SiO emission 
arises in only the cool, dense component. The corresponding relative 
abundances of SiO, $X$(SiO), are 25\% above the beam averaged values given 
in Table\,\ref{col}. $X$(SiO) lies within the range allowed by the SiO 
line ratios (Fig.\,\ref{lvg1}) and is consistent with an H$_2$ density 
of typically (1~--~4)\,10$^4$\,\percc. The beam filling factor  
$f_{\rm b}$ in this dense component must be the same for both the SiO 
and the part of the \CEIO\ emission that arises in this component. 

We can estimate the main beam brightness temperature of the dense,
\CEIO\ emitting gas to be 0.75 of the total \CEIO\ intensity. Using the 
beam filling factor determined for the dense gas from SiO lines, 
the brightness temperature of the \CEIO (2 $\to$ 1) line is 
$T_{\rm B}^{\rm C^{18}O}$(75\%)\,$\approx$\,0.75\,\TMB\,$/f_{\rm b}$. 
This simple relation can be used since the \CEIO ($2\to 1$) line has an 
optical depth $\leq 1$. If the solution is self-consistent, 
the H$_2$ density indicated by this value of \TB\ and the observed line 
ratios must agree with the density derived for SiO. To within a factor
of 2, we find that this is indeed the case. Thus, it is possible 
that the SiO emission originates in gas having \numd\,$\sim 
(1-4) 10^4$\,\percc. This gas must be relatively
cool, which is a surprising result in the light of the clear association 
of the SiO emission with hot shocked gas in the Galactic disk. The 
intrinsic optical depth $\tau_0$ of the SiO($2\to 1$) then ranges from 4 -- 8. 

Can the SiO emission be associated with the hot, thin gas component? 
To check this, we used the same procedure as before: From the assumption 
that the SiO emission arises in $\sim 25$\% of the gas, we get a value 
for $X$(SiO) and a corresponding H$_2$ density. The beam filling factor 
$f_{\rm b}$ for SiO and, consequently, for the part of \CEIO\ originating 
in the same gas does not change. Now, 
$T_{\rm B}^{\rm C^{18}O}$(25\%)\,$\approx 0.25\,$\TMB\,$f_{\rm b}$. 
The additional requirement that this gas should be hot yields a H$_2$ 
density that is, for many clouds, lower by an order of magnitude than 
what is derived for the same component from SiO. Thus, this is not a 
consistent solution. If the SiO arises in a still smaller portion of the 
total gas, the discrepancy becomes even worse. Note that we do not claim 
that the hot, thin component is devoid of SiO. We consider it likely that
the SiO abundance in this component is similar to what we determine for
the dense component. However, the SiO excitation in this component is 
extremely subthermal, leading (1) to lower line intensities than those we 
observe (specifically, no emission from the (5$\to$4) transition) and (2)
to different line intensity ratios.  

Finally, we can rule out the presence of a hot, high density 
(\numd\,$\geq 10^4$\,\percc) component as the origin of the SiO
emission: We know from the line ratios \RMSIO\ and \RISOSIO\ that 
$X$(SiO) must decrease with increasing density, independent of 
\TKIN . Thus, a hot, high density component containing the 
bulk of the SiO would have to contain a significant fraction of the 
total H$_2$ column density and mass. Such a component should be visible in
\CEIO . Thus we conclude that it does not exist in these clouds. 

Therefore the only consistent scenario that
accounts for both the SiO and \CEIO\ results requires that most of the
SiO emission arises from a {\em cool} (20\,K -- 30\,K) gas component 
with \numd\ $\sim 10^4$\,\percc . 

So far, we have {\em assumed that the \CEIO\ abundance in
all components considered is known and constant.} If this is not the 
case, i.e.\ if \CEIO\ is {\em not} a reliable H$_2$ column density tracer,
the line of argumentation given above cannot be maintained. In this case,
it becomes possible to claim that the SiO emission arises in a hot, dense 
component not seen in \CEIO . This is, however, only possible if \CEIO\
is selectively underabundant in the component where the SiO abundance is
high. Since selective \CEIO\ dissociation due to lack of shielding is
efficient only in diffuse, low density gas (if the UV field is strong
enough), it is difficult to find a scenario causing \CEIO\ to be
underabundant in dense, warm gas. 

\subsection{A detailed investigation of two sources}

\begin{figure*}
\psfig{file=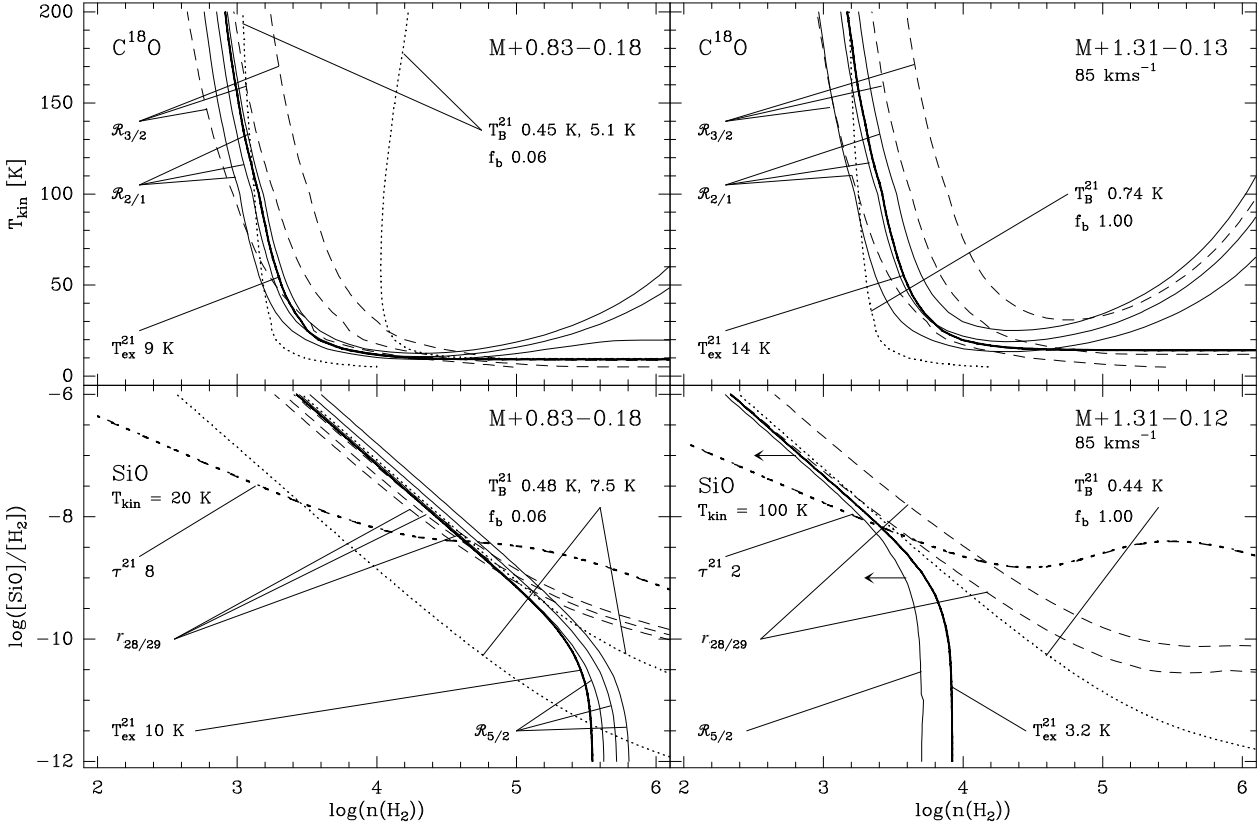,bbllx=25pt,bblly=40pt,bburx=560pt,bbury=680pt,height=12.2cm,angle=-90}
\caption{ \label{lvg} LVG Model calculations for two individual sources,
demonstrating the range of physical conditions encountered. The observed line
intensity ratios are plotted including their 1\,$\sigma$ uncertainties.
The \TB - curves are the observed \TMB\ and the best fit temperature derived
from the line ratios. The coresponding beam filling factor $f_{\rm b}$ is
also given. For M+1.31--0.13, the observed and the best fit temperature are
the same, thus $f_{\rm b} = 1$. $\tau^{21}$ is the peak optical depth 
of the $^{28}$SiO line (not beam averaged). }
\end{figure*}

\paragraph{M+0.83--0.18}
In Fig.\,\ref{lvg} (left panels), we illustrate the procedure of 
deriving the physical cloud parameters described above for the 
source M+0.83--0.18. This is representative of the type of cloud where 
SiO emission arises in a cool, moderately dense component. The SiO
line ratios in this source (lower left panel in Fig.\,\ref{lvg}) are well
fit by \TEX\,$= 10$\,K. Then, $X$(SiO) is 3.9\,10$^{-9}$ if SiO arises 
from 75\% of the total gas. The observed \TMB\ is 0.48\,K, while a
\TB\ of 7.5\,K is required to fit the range of \numd\ and $X$(SiO).
This results in the lowest value for $f_{\rm b}$ of all sources, 0.06, and
the highest optical depth $\tau_0$, 8. Applying the same $f_{\rm b}$ to 
the \CEIO\ data (upper left panel), we have 
\TB (\CEIO)\,$=$\,0.75\,\TMB /0.06\,=\,5.1\,K.
If we take the resulting \numd\ from the SiO analysis, 4\,10$^4$\,\percc ,
and transfer this to \CEIO , we find that \TKIN\ is only $\sim 12\,$K. 
If we require \TKIN\ to be identical to the value we have measured for 
the cool component from NH$_3$, 21\,K (H\"uttemeister et al.\ 1993b), 
the density obtained for \CEIO\ is 2\,10$^4$\,\percc . Considering all 
the assumptions, there is satisfactory agreement. Could the results be
consistent with the `high temperature scenario' described in the last section? 
We find that \numd\ required from an analysis of the SiO data is 
$\sim 3\,10^4$\,\percc, while the \CEIO\ data at the appropriate 
\TB\ yield $\sim 3\,10^3$\,\percc , clearly a far less consistent 
solution. 

\paragraph{M+1.31--0.13}
Now we analyze the one source which is very different from all the others
in our sample: M+1.31--0.13 (right panels of Fig.\,\ref{lvg}). This is
the only source where the ($5\to 4$) transition of $^{28}$SiO was not
detected, even though the ($2\to 1$) line is strong. It also is the one
source where \RISOSIO\ approaches the terrestrial isotopic ratio of 20,
indicating that $\tau_0$ for $^{28}$SiO($2\to 1$) cannot be large. M+1.31--0.13
has two velocity components. We analyze the (slightly) more intense  
\vlsr\,$\sim 85$\,\kms\ component, but the \vlsr\,$\sim 35$\,\kms\
component shows the same characteristics. 

From the SiO line ratios, we derive an extremely low excitation
temperature, 3.2\,K. The large column density for SiO
depends strongly on the exact value of \TEX . Therefore, we can 
give only an estimate of $\bar N^{\rm SiO}$ in
Table\,\ref{col}.  The corresponding $X$(SiO) is $\sim 10^{-8}$, 
by far the highest in the sample. In addition, for this source 
we find \TMB\,$=$\,\TB , i.e.\ $f_{\rm b} = 1$. As expected, this 
results in a lower $\tau_0$, $\sim 2$, in the $^{28}$SiO($2\to 1$) 
line. For the H$_2$ density, we find \numd\,$\approx 2\,10^3$\,\percc . 
This is a factor of $\sim 10$ lower than for all other sources. 

Since our calculations give $\tau_0 \sim 2$ and $f_{\rm b} = 1$ for 
SiO, we conclude that in this source SiO traces {\em all} the gas, as 
does \CEIO . Hence, $f_{\rm b} = 1$ and \TB\,$=$\TMB\ is also valid for 
\CEIO . The resulting \numd\ from \CEIO\ is $2\,10^3$\,\percc , 
identical to the result obtained from SiO. As expected, \TKIN\ is 
high, $>\,100$\,K.  Thus, for this one source, we have found 
a fully self-consistent solution of an entirely different type: The 
SiO arises in thin, hot gas and has an abundance that is far above 
average, even for the clouds that are rich in SiO. 

Is there evidence that M+1.31--0.13 lacks the dense and cool
gas component? Remarkably, our NH$_3$ data show that this cloud is {\em 
the only one in the entire sample} in which the $(J,K)=(4,4)$ inversion
transition at 200\,K above ground is more intense than the (1,1) transition
at 23\,K above ground. Thus, in this cloud a cool, dense gas component
containing the bulk of the gas is not present. Most, perhaps all
of the gas is hot and thin. Relating this to the high SiO abundance, it is
likely that SiO does indeed form at high temperatures in the Galactic 
center region. In M+1.31--0.13, the SiO formation process is either 
still ongoing or has occured very recently. The SiO has not yet had 
time to recondense onto dust grains, and the cloud itself has not had 
time to form dense cool cores.

\section{Discussion}

\subsection{Cloud properties as a tracer of large scale dynamics}

We will now explore how the relative SiO abundances may be related to the 
position of the cloud in the large scale Galactic center environment.

\begin{figure}
\psfig{file=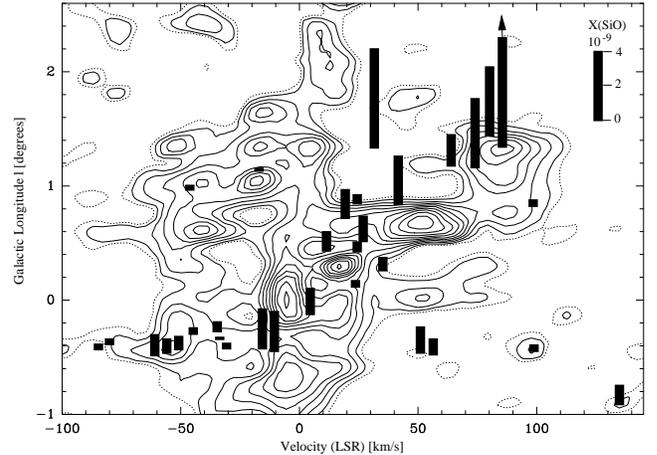,bbllx=0pt,bblly=55pt,bburx=530pt,bbury=700pt,height=6.7cm,angle=-90}
\caption{\label{nl} The beam averaged SiO abundance plotted on a longitude-
velocity diagram of the large scale distribution of \CEIO , adapted from
Paper\,I (spatial resolution 9$'$). }
\end{figure}

In Fig.\,\ref{nl} we plot the the beam averaged SiO abundance, 
$X^{\rm SiO}$, on a longitude-velocity diagram of the large 
scale distribution of \CEIO . While any estimates of line-of-sight 
locations in the Galactic center are rather uncertain, this plot 
gives a measure of location.   

There is a clear trend in the large-scale distribution of
SiO abundances. For $-0\ffd 35 \leq l \leq 0\ffd 6$,
between the location of the Sgr\,C and Sgr\,B2 regions, we
find a pronounced lack of very large $X$(SiO). Most clouds
with very high abundances, including the exceptional source M+1.31--0.13, 
are located at $l > 0\ffd 8$. The `Clump\,2' Region is not 
included in this plot, since, at $l \approx 3^{\circ}$, this is not really 
part of the continuous molecular bulge or, in the notation of Morris 
\& Serabyn (1996), the Central Molecular Zone (CMZ), although it shares 
many characteristics of the Galactic center gas. $X^{\rm SiO}$ in this region 
ranges from 1\,10$^{-9}$ to 2\,10$^{-9}$, typical `Galactic center' values 
that are high compared to the disk but not as extreme as what is found 
at $0\ffd 8 < l < 1\ffd 5$.

The detection of an `SiO hole' at  $-0\ffd 35 \leq l \leq 0\ffd 6$ is 
confirmed by an extension of the survey of Mart\'{\i}n-Pintado et al.\ 
(1997) (unpublished data). 

In recent years, the large scale dynamics of the gas in the CMZ, 
characterized by large non-circular motions and a distinct 
`parallelogram' shape of $lv$-diagrams based on $^{12}$CO and $^{13}$CO
(e.g.\ Bally et al.\ 1988), have been explained in terms of a model 
involving a rotating (stellar) bar with corotation at 2.4\,kpc and 
oriented at an angle of $\sim 20^{\circ}$ with the line of sight  to the
Galactic center (see Morris \& Serabyn 1996 and references therein).  Within
such a potential, there exist two categories of closed elliptical orbits,
called $x_1$ and $x_2$-type orbits. Inside a cusped orbit, the
$x_1$-orbits, which are elongated along the bar axis, become self-intersecting. 
Clouds on these orbits encounter a shock and within a dynamical time plunge
into $x_2$-orbits that lie considerably deeper within the potential and
mimic circular orbits (Mulder \& Liem 1986, Athanassoula 1988, 1992).  
This encounter  breaks the flow along the cusped orbit into a spray that
spreads  out into the interior of the orbit (Binney et al.\ 1991, 
Athanassoula 1992, Jenkins \& Binney 1994). 

According to this model, Sgr\,B2 and Sgr\,C can be interpreted to be at the
locations of the intersections of the $x_1$- and the $x_2$-orbits. The dense,
virialized clouds on the $x_2$-orbits, inside of Sgr\,B2 and Sgr\,C,
are more likely to form stars, while the less dense clouds on the  $x_1$
orbits are too disrupted by shocks for efficient star formation. Generally, 
this is supported by the fact that stars of young and intermediate age are
restricted to the area inside of Sgr\,B2 and Sgr\,C, suggesting sustained
star formation in this region (Serabyn \& Morris 1996).  In Sgr\,B2, the
ongoing star formation, the large amount of molecular gas, the 
existence of a very hot molecular component (H\"uttemeister et al.\ 1993a, 
1995, Flower et al.\ 1995) and the high optical
depth of the $^{12}$CO emission (Paper\,II) also provide
ample evidence for this picture. The Sgr\,C complex is a much more 
quiescent region, but it contains a large H\,{\sc II} region and GMCs.

Because the sprayed gas crashes into  material that is still on 
$x_1$-orbits, collision regions are expected along the acceleration
part of the cusped orbits.  Due to
perspective, these main collision areas should be at higher positive
longitudes than Sgr\,B2 and between Sgr\,A and Sgr\,C. While the 
situation at negative $l$ is confused (the gas on $x_2$ orbits and
the collision regions are expected at similar Galactic longitudes),
at positive $l$, especially the `$1\ffd 5$--complex' (Bally et 
al.\ 1988) shows all characteristics expected from a collision region
(see the discussion in Paper\,II). 

Inspecting Fig.\,\ref{nl} with this picture in mind, we find remarkable
support from our data. The exceptional cloud M+1.31--0.13 and
three other sources with high SiO abundances are located within
the 1.5$^{\circ}$--complex, exactly where strong shocks are expected.
Generally, the high SiO abundances at $l > 0\ffd 8$ can be 
identified with a region in which the probability of gas encountering 
shocks is high. Thus, for the first time, we can tie the physical
and chemical parameters of Galactic center molecular peaks to the 
large scale dynamics of the region. 

Since CS traces all dense gas, not just the part that has been subjected
to shocks, we do not expect a similar `zone of avoidance'.
Indeed, a plot similar to Fig\,\ref{nl} comparing the CS abundance in 
these clouds (data presented in H\"uttemeister 1993), to the total H$_2$
column density does not show such an effect.  

Of course, the correspondence is not absolute. Not every cloud at 
$l > 0\ffd 8$ {\em must} experience strong shocks. Note, however, that 
the clouds with very low SiO abundances in this range of $l$ have 
negative velocities and might not be located in the collision region. 
Also, there are certainly strong shocks occuring in the 
Galactic center that are unrelated to large scale dynamics. Mart\'{\i}n-Pintado 
et al.\ (1997) explain the `SiO clouds' they find by shocks of a variety
of origins: Interaction with SNRs close to Sgr\,A, interaction
with non-thermal filaments in the radio arc and cloud-cloud collisions
(as expected from large scale dynamics) or expanding bubbles in the
vicinity of Sgr\,B2. Their map extends from $-0\ffd 2$ to $+0\ffd 8$.
Therefore, they have missed the systematic signature of large scale 
effects present at higher positive longitudes. 

Another mechanism explaining the preferential occurence of shocks in the 
$1\ffd 5$--complex might be fossil superbubbles, remnants of a phase 
of Sgr\,B2 type star formation activity. The expansion of 
superbubbles would offer a natural explanation of the large extent of the 
CO emission to positive Galactic latitudes perpendicular to the Glactic 
plane seen in this region. However, while the bar model as outlined above is 
two-dimensional and thus does not naturally produce vertical structure, 
sprayed gas colliding with gas on $x_1$-orbits might give rise to turbulence 
also pushing gas out of the plane. It is noteworthy that a $^{12}$CO map 
(see, e.g., Paper\,II and Bitran et al.\ 1997) shows the molecular gas 
extending toward negative $b$ at the southern edge of the CMZ. 
Thus, it seems possible that the CMZ is warped or susceptible to 
instabilities close to its edges (Morris \& Serabyn 1996 and references 
therein). 

The `bar model' is not the only possibility 
to explain the large scale dynamics of the Galactic center region. 
von Linden et al.\ (1993) suggest that an accretion disk can also 
reproduce the basic structure seen in position-velocity plots. While 
this is certainly true for the inner region ($x_2$-orbits are almost 
indistinguishable from circular orbits) it is not clear whether such
a scenario can also explain the gas distribution and 
the different chemical properties in the entire CMZ. Based on the data 
presented in Papers I and II and this work, we will examine 
the questions related to large scale structure and dynamics closely in 
a forthcoming paper (von Linden et al.\ 1998).  

\subsection{The origin of SiO}

Even though the SiO emission we observe is mostly 
associated with cool gas,  we consider it likely that it forms 
under conditions of high kinetic temperature and grain erosion by shocks
(Mart\'{\i}n-Pintado et al.\ 1992). Both the distribution discussed in the 
previous section and the existence of the hot cloud core 
M+1.31--0.13 can be taken as evidence for this. From the fact that only
one cloud in our sample, namely M+1.31--0.13, shows hot gas emitting in
SiO lines, we can conclude that shocked gas forms dense, cooler cores 
fairly fast. Extrapolating from the analysis in Hollenbach (1988), we find a
very short cooling time scale ($\ll 10^6$\,yr) for gas at a density
of 10$^4$\,\percc . A competing  process is the condensation of gas phase 
SiO onto grain mantles. We estimate this `freeze out' time scale to be of 
order $10^6$\,yr or slightly less (Rohlfs \&  Wilson 1996). Thus,
gas with a high gas phase relative abundance of SiO has been cool for most
of its lifetime.  

\begin{figure}
\psfig{file=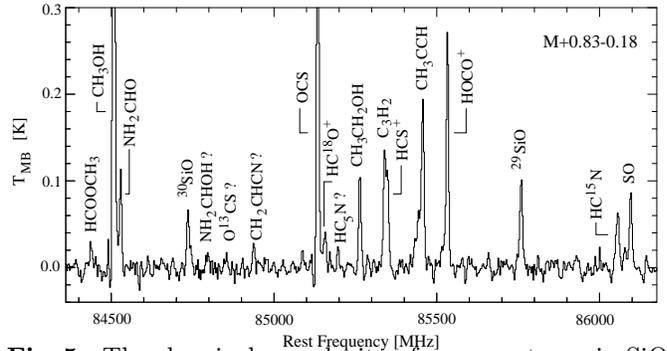,bbllx=170pt,bblly=55pt,bburx=520pt,bbury=600pt,height=4.2cm,angle=-90}
\caption{\label{complex} The chemical complexity of sources strong
in SiO is demonstrated by our spectrum of M+0.83--0.18, covering
$\sim 1.8$\,GHz and including both $^{29}$SiO and $^{30}$SiO. The spectrum 
shown is a composite of two single SEST low frequency resolution spectra.}
\end{figure}

After more than 10$^6$ years, the gas slowly loses the chemical memory
of having been shocked. From the typical radial velocities of
the clouds, we estimate that, during this time, these can move 
$\sim 50$\,pc. If the very SiO-rich clouds are preferentially located on
$x_1$-orbits along the bar, these will remain sufficiently close in longitude 
to the region where they encountered the shock to be recognised as a distinct
population. 

Since we find some SiO in all cloud cores, additional formation of gas phase
SiO is required, probably by local turbulence and/or cloud-cloud collisions 
causing some shocks. This is a general characteristic of the cores of 
Galactic center GMCs. Ongoing local shock activity is likely to be also
necessary as the main heating mechanism for the hot gas component seen in
all clouds (see Flower et al.\ 1995 and the discussion in H\"uttemeister 
et al.\ 1993).

It is known that gas in post-shock regions, away from chemical equilibrium,
is characterized by abundant complex molecules (e.g.\
Brown et al.\ 1988). This is exactly what can be observed toward the 
positions within the CMZ that are strong in SiO. We show just one example in
Fig.\,\ref{complex}. In the Galactic disk, such complex spectra are
typical for very confined regions, while in the CMZ they seem to be
ubiquitous. This chemical complexity can only be maintained for 
$\sim 10^6$ years and thus requires frequent shocks. These may also be typical
for the starburst environment (see Henkel et al.\ 1987, Mauersberger \& Henkel
(1991), Mauersberger et al.\ 1991, H\"uttemeister et al.\ 1997 for the 
case of NGC\,253). On larger timescales, an equilibrium state is
reached. This consists of a gas phase component of mostly diatomic 
molecules (Herbst \& Leung 1989). 

Most time-dependent chemical models have been calculated for either cold,
quiescent clouds (e.g.\ Bergin et al.\ 1995) or 'classical' collapsing, star-forming hot cores (e.g.\ Brown et al.\ 1988, Caselli et al.\ 1993). 
Thus, detailed, model-supported recommendations of the molecules that should be
used in future work to trace the chemical evolution of the the Galactic center 
cores cannot be given. CH$_3$OH and SO$_2$ appaer, however, to be promising
species. The former is a high temperature, high density tracer requiring 
\TKIN $>$ 70\,K to evaporate from dust grains. Its wealth of emission lines
allows density and temperature determinations from multi-level studies. 
Observations in external galaxies show that its abundance can differ widely
even in the central regions of galaxies (H\"uttemeister et al.\ 1997). 
SO$_2$ is associated with shock chemistry and grain destruction. 
The abundance of the two molecules need not be correlated, as is shown by 
an evolutionary study of cores in the W3 region (Helmich et al.\ 1994). Both
molecules have lines that are strong enough to allow large scale mapping in 
the Galactic center region, and differences in their distribution would give
further insight into the predominant processes operating on the Galactic
center cores. 

\section{Conclusions}

Based on our measurements and analysis of 33 molecular density peaks 
in the Galactic center in the $J = 1\to 0, 2\to 1$  and $3\to 2$ 
transitions of \CEIO , \ the $J = 2\to 1$ and $5\to 4$ transitions 
of $^{28}$SiO and the $J = 2\to 1$ transitions of $^{29}$SiO 
and $^{30}$SiO, we find:

\begin{enumerate}
\item All sources are easily detected in all transitions of \CEIO\ 
searched for and in $^{28}$SiO($2\to 1$), demonstrating that the  
properties of molecular peaks in the Galactic center region are 
markedly different from the Galactic disk, where thermal SiO 
emission is confined to very small regions in the vicinity of 
outflows associated with star formation. Local gas on the 
line-of-sight, distinguished from Galactic center gas by its 
narrow lines, is not detected in SiO.

\item The rare isotopomers $^{29}$SiO and $^{30}$SiO are detected in
all 12 studied sources. The $J=5\to 4$ transition of the main isotopomers 
is seen in 7 out of 8 sources. The line intensity ratio 
of $^{29}$SiO and $^{30}$SiO shows that the terrestrial 
isotope ratio, $^{29}$Si/$^{30}$Si\,$=$\,1.5, holds for the Galactic
center region.

\item From LVG model calculations applied to the \CEIO\ line intensity 
ratios \RZWCO\ and \RDRCO , cool (\TKIN\,$\sim 20 -30$\,K) gas toward 
the molecular peaks has moderately high densities (\numd\,$\sim 
(1 - 4)\,10^4$\,\percc), while high kinetic temperatures of $> 100$\,K
correspond to H$_2$ densities that are an order of magnitude
lower. This is contrary to what is found in the disk, where the 
cores of GMCs are usually hot, and is an indication that the
`cool cores' in `typical' Galactic center GMCs, away from Sgr\,A and
Sgr\,B2, are not presently forming high mass stars.

\item Combining the results of LVG models for \CEIO\ and SiO, and 
using \CEIO\ as a tracer of total H$_2$ column density, 
a beam averaged SiO abundance is derived for all clouds. This varies 
significantly from source to source, ranging from $\sim 0.3\,10^{-9}$
to $> 5\,10^{-9}$. The $^{28}$SiO($2\to 1$) transitions are optically
thick, with $\tau_0$ ranging from $2 - 8$. 

\item Including information on the temperature structure of the 
clouds from NH$_3$, it is shown that for most clouds a self-consistent
solution accounting for the properties of \CEIO\ and SiO is only
possible if the bulk of the SiO emission arises in the cool, dense 
gas component. 

\item One source, M+1.31$-$0.13, is different from all the others:
SiO arises in the hot, thin gas component. Since this is also the 
cloud with the highest SiO abundance in the entire sample, the SiO 
formation process, probably grain erosion by shocks, is likely to be 
still ongoing in this source.

\item The SiO abundance in individual clouds is related to the large
scale gas dynamics in the Galactic center region. The highest 
abundances are found at Galactic longitudes of $l > 0.8^{\circ}$,
which can be identified with `collision regions' likely to encounter
shocks in terms of the bar model of gas dynamics. 

\item As in the disk, SiO in the Galactic center region is likely 
to originate in shocks. Since the hot, thin post-shock gas forms 
dense, cool cores faster than SiO recondenses to dust grains, the
SiO rich gas is cool for most of its lifetime. On a timescale
of $\sim 10^6$\,yr, the SiO molecules freeze out on grain mantles 
and the clouds lose their chemical memory of the shock.
\end{enumerate}

\acknowledgements{We thank the staff of the telescopes,
especially L.\ Knee and L.-\AA.\ Nyman (SEST), and the SMTO/HHT
staff for their help in taking the observations. GD gratefully 
acknowledges fiancial support from the Particle Physics and Astronomy
Research Council (PPARC) (UK), for the observing run at JCMT. RM 
was supported by a Heisenberg fellowship by the Deutsche 
Forschungsgemeinschaft. JM-P has been partially supported by the 
Spanish DGES under grant number PB96-0104. }

\end{document}